\documentclass[aip,rsi,reprint,groupedaddress]{revtex4-1}
\usepackage{graphicx}
\usepackage{times}
\usepackage{color}
\usepackage{amsmath}
\usepackage[english]{babel}

\newcommand{\bn}{\begin{equation}}
\newcommand{\en}{\end{equation}}
\newcommand{\Nat}{N_{\rm at}}
\newcommand{\CS}{\rm CS}

\newcommand{\Deff}{\Delta_\mathrm{eff}}
\newcommand{\rme}{\mathrm{e}}

\begin{document}
\title{Non-destructive Faraday imaging of dynamically controlled ultracold atoms}
\author{Miroslav Gajdacz}
\author{Poul L. Pedersen}
\author{Troels M\o{}rch}
\author{Andrew J. Hilliard}
\author{Jan Arlt}
\author{Jacob F. Sherson}\email[]{sherson@phys.au.dk}
\affiliation{Danish National Research Foundation Center for Quantum Optics, Institut for Fysik og Astronomi, Aarhus Universitet, Ny Munkegade 120, 8000 Aarhus C, Denmark.}

\date{\today}

\begin{abstract}
We describe an easily implementable method for non-destructive measurements of ultracold atomic clouds based on dark field imaging of spatially resolved Faraday rotation. The signal-to-noise ratio is analyzed theoretically and, in the absence of experimental imperfections, the sensitivity limit is found to be identical to other conventional dispersive imaging techniques.
The dependence on laser detuning, atomic density and temperature is characterized in a detailed comparison with theory. Due to  low destructiveness, spatially resolved images of the same cloud can be acquired up to 2000 times. The technique is applied to avoid the effect of shot-to-shot fluctuations in atom number calibration, to demonstrate single-run vector magnetic field imaging and single-run spatial imaging of the system's dynamic behavior. This demonstrates that the method is a useful tool for the characterization of static and dynamically changing properties of ultracold atomic clouds.
\end{abstract}

\maketitle

\section{Introduction}

One of the most fundamental features of quantum mechanics is the appearance of intrinsically non-classical effects induced by measurements. An important class of measurements  for implementing various quantum technologies is that of quantum-non-demolition (QND) measurements\cite{Braginsky1996}, which have been realized in ionic\cite{Hume2007}, superconducting\cite{Lupascu2007}, optical\cite{Grangier1998} and microwave\cite{Guerlin2007,Johnson2010a}  systems, and in atomic ensembles~\cite{Hammerer2010}.

Spatially resolved non-destructive imaging of atomic ensembles has  recently been realized using diffractive methods~\cite{turner2005} and    partial-transfer absorption imaging~\cite{Ramanathan2012}, but typically it is achieved  by dispersive methods\cite{Ketterle_review}. Dispersive methods rely on an atomic sample imparting a phase shift on imaging light by either the scalar or  vector part of the interaction Hamiltonian. Methods relying on the scalar part  include dark field scalar imaging (DFSI)~\cite{Andrews1996} and phase contrast imaging (PCI)~\cite{Andrews1997}. Methods based on the vector part use the anisotropic response of different magnetic substates - commonly known as the Faraday  effect - and have been implemented in two variants analogous to the mentioned scalar methods: dark field Faraday imaging  (DFFI)~\cite{Bradley1997} and dual port Faraday imaging  (DPFI)~\cite{Kaminski2012}. The vector nature of the Faraday interaction has previously been used to demonstrate entanglement~\cite{Julsgaard2001}, quantum memory~\cite{Julsgaard2004}, and quantum teleportation~\cite{Sherson2006a} in room temperature atomic ensembles. In the regime of cold atoms, this approach has yielded spectacular results including spin squeezing~\cite{Takano2009}, magnetometry~\cite{Vengalattore2007,Fatemi2010,Smith2011,Koschorreck2011}, and the observation of many-body dynamics~\cite{Liu2009}. Although the vector methods have  been used for the detection of optical densities~\cite{Bradley1997,Kaminski2012}, they are often considered to be inferior to scalar methods and neglected in  signal-to-noise ratio (SNR) analyses of various protocols~\cite{PhysRevA.67.043609,Ramanathan2012}.

Here, we present a detailed experimental characterization of DFFI, which can be realized by inserting a single polarizer in a standard absorption imaging set-up and is thus considerably simpler to implement than the other dispersive techniques. We show that DFFI allows for precise measurements over a wide range of atomic densities and temperatures and evaluate the detuning dependence of the method and its destructive effects on the atomic sample. 

We also present a detailed comparison of the four dispersive imaging techniques noted above. The signal in dark field techniques such as DFFI and DFSI scales quadratically with the acquired phase shift for small angles, whereas PCI and DPFI scale linearly such that the latter are typically  considered to be superior~\cite{Ketterle_review}. However, when assessing the quality of a detection method, the decisive parameter is the signal-to-noise ratio. A comparison of the achievable SNR  in these four dispersive techniques shows that, up to a  factor of the order one,  all yield  the same sensitivity. This unexpected result arises from the fact that the fundamental source of noise is the shot noise of light; in the dark field techniques, less light hits the detector, leading to a corresponding reduction in the noise.

\begin{figure}[t]
	\includegraphics[width=8.5cm]{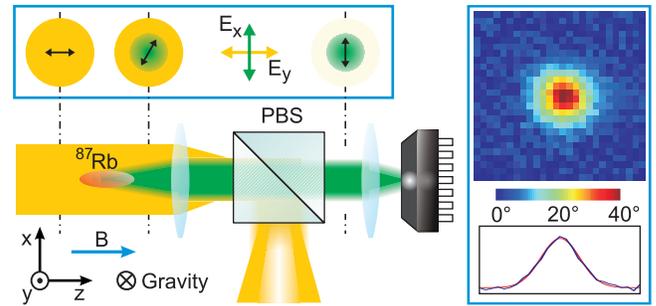}	
	\caption{Sketch of the Faraday imaging system and a resulting image (right) at $T=1.5~\mu$K and $N=10^6$ atoms. The polarization of the light is indicated (top left) by displaying cross sections of the imaging beam obtained from a rotation around the dash-dot lines.}
  \label{fig:imaging_setup}
\end{figure}

This paper is structured as follows. We begin with a theoretical analysis of the DFFI method and its experimental characterization. We then turn to applications of the method by first  demonstrating  multiple probing of a cloud of ultracold atoms with up to 2000 images per experimental cycle. We present three diagnostic applications: the mitigation of shot-to-shot atom number fluctuations by rescaling absorption measurements with Faraday images,   the determination of ambient vector magnetic fields, and the measurement of trapping frequencies that are varied within a single experimental run. The second half of the paper provides a detailed comparison of  dispersive imaging techniques, presenting  the signal-to-noise ratio for each of the four methods and a discussion of our findings.

\section{ Analysis of the DFFI method}

\subsection{Theoretical model}

Faraday rotation can be described using a model of the dispersive atom-light interaction~\cite{Hammerer2010}.
The effective scattering Hamiltonian $\hat{H}_\mathrm{eff} = \hat{H}^{(\mathrm{scal})} + \hat{H}^{(\mathrm{vec})}$
consists of a scalar and a vector part
\begin{eqnarray}
   \label{scal_ham}
   \hat{H}^{(\mathrm{scal})} &=& \frac{1}{3}g \sum_{f^{\prime}} \frac{\alpha^{(\mathrm{scal})}_{f,f^{\prime}}}{\Delta_{f,f^{\prime}}}
   \hat{N}_{\rm at} \hat{N}_{\rm ph}, \\
   \hat{H}^{(\mathrm{vec})} &=& \frac{1}{2} g \sum_{f^{\prime}} \frac{\alpha^{(\mathrm{vec})}_{f,f^{\prime}}}{\Delta_{f,f^{\prime}}} \hat{F}_z
   \left( \hat{N}_+ - \hat{N}_-  \right),
   \label{inter_ham}
\end{eqnarray}
where $\hat{N}_{\rm at}$ and $\hat{N}_{\rm ph}$ are atom and photon number operators respectively, $\hat{F}_z$ is the $z$-component of the collective atomic angular momentum (defined by the direction of light propagation) and $\hat{N}_\pm$ are photon number operators for the two circular polarizations.
The scalar and vector polarizabilities $\alpha^{(\mathrm{scal})}_{f,f^{\prime}}$ and $\alpha^{(\mathrm{vec})}_{f,f^{\prime}}$
characterize a given atomic transition and \mbox{$\Delta_{f,f^{\prime}} = \omega - \omega_{f,f^{\prime}}$} is the detuning of the
light frequency $\omega$ from the atomic transition. Finally, the field factor is \mbox{$g=\omega/(2 \epsilon_0 V)$}, where $V$ is the atom-light interaction volume.

Provided the atom number $\Nat$ is large and all atoms are in the same internal state with average $z$-axis projection
of the angular momentum $\langle \hat{f}_z \rangle$, we can treat the collective angular momentum classically
and use ${\hat{F}_z  = \Nat \langle \hat{f}_z \rangle}$. To account for the spatial variation of the density, we substitute $\Nat/V  \rightarrow \rho({\bf r})$.
Denoting the eigenvalues of the number operators $\hat{N}_\pm$ by $E_\pm$, the Hamiltonian \eqref{scal_ham} induces the scalar phase shift $\theta_\mathrm{S} = \int \frac{1}{2}(E_+ + E_-)dt/\hbar$  used in, e.g., PCI. The vector term \eqref{inter_ham} gives rise to a differential phase shift
of the two circular components $\theta_\mathrm{F} = \int \frac{1}{2}(E_+ - E_-)dt/\hbar$. This effect ---known as  Faraday rotation--- causes a rotation of the polarization plane of initially linearly polarized light.

In our experiments, atoms are prepared in the ${|f=2, m_f = 2 \rangle}$ state of $^{87}$Rb, and for imaging wavelengths close to the D2 transition, the spatially resolved Faraday angle is given by
\bn
\label{far_rot}
\theta_\mathrm{F}(x,y) = \frac{ \langle \hat{f}_z \rangle \Gamma \lambda^2 }{16\pi \Deff} \int \rho({\bf r}) dz=c_F(\Deff) \tilde{\rho}(x,y),
\en
where $\Gamma$ is the natural linewidth, $\lambda$ is the wavelength of the imaging light and the effective detuning is given by
\bn
\label{ef_det}
\frac{1}{\Deff} = \frac{1}{20} \left( \frac{28}{\Delta_{2,3}}- \frac{5}{\Delta_{2,2}} - \frac{3}{\Delta_{2,1}} \right).
\en
For further analysis, we represent the spatially dependent angle of polarization as a product of a Faraday coefficient $c_F(\Deff)$ and the column density of the sample $\tilde{\rho}(x,y)$.
	
Figure~\ref{fig:imaging_setup}  shows a schematic of the experimental setup used to measure this angle of polarization. When a beam of linearly polarized light impinges on a cloud of magnetically oriented atoms, a spatial rotation pattern is imprinted on the beam.
The polarization pattern is collimated and the two linear components are subsequently separated on a polarizing beam splitter (PBS).
The polarization of the imaging beam is chosen such that its transmission through the PBS is minimized in the absence of atoms. A second lens then forms an image in the detection system, which contains a mask to allow for partial  readout of the camera and thereby high  frame rates. We have realized frame rates up to 2~kHz, which is somewhat slower than the fastest implementations\cite{Higbie2005}. However, this allows us to configure the camera for continuous acquisition, which is crucial for the applications discussed below. 
We employ an Electron Multiplying Charge Coupled Device (EMCCD) camera; this enables low light intensity imaging and hence reduced measurement destructiveness. The EM gain  is crucial for repeated probing and feedback experiments but comes at the expense of an amplification of the shot noise---noise arising from the quantum nature of light---by a factor of $\sqrt{2}$, which can be a severe limitation in applications with low signal-to-noise ratio~\cite{Robbins_IEEE_50_1227_2003,Hynecek2003}.

The reconstruction of  the rotation angle requires knowledge of the intensities of the incoming and the rotated light. In principle, this can be achieved by measuring the intensity in both ports of the PBS~\cite{Kaminski2012}. In our realization, however, we avoid this   by making use of the experimental imperfection of the PBS, which leads to a finite transmission of non-rotated light. Thus, images without atoms can be used to determine the incoming light intensity and compensate for  beam profile inhomogeneity. The transmitted light intensity is given by
\bn
\label{eqn:thetaF}
   I(\theta_\mathrm{F}) = I_0 \frac{\sin^2 \theta_\mathrm{F}  + \CS \cos^2 \theta_\mathrm{F}}{1 + \CS} ,
\en
where $I_0$ is the incident intensity. The cube suppression
$\CS = I(0)/I(\pi/2) = 1.5 \times 10^{-3}$
is the ratio of minimum to maximum light intensity transmitted through the PBS for a manually scanned polarization angle.
The Faraday rotation angle can then be obtained from
\bn
\label{sintheta}
  \sin^2  \theta_\mathrm{F} = \left(\frac{I(\theta_\mathrm{F})}{I(0)} - 1 \right) \left( \frac{\CS}{1 - \CS} \right).
\en
An absolute light intensity calibration is therefore not required to evaluate the rotation angle as long as the \mbox{EMCCD} camera has a linear response.

\subsection{Experimental characterization}

The Faraday imaging technique requires light detuned from atomic resonance by $\sim100~\Gamma$. This light is produced by an extended cavity diode laser locked to a reference laser via a tunable offset lock~\cite{Engler1999}. Since the ${f = 2 \rightarrow f^\prime = 3}$ transition has the highest oscillator strength (Eq.~\eqref{ef_det}), we use \mbox{$\Delta_{2,3} \equiv \Delta$} as a measure of the laser detuning. This setup allows us to lock the laser in the range \mbox{$\Delta = (-1.5, 1.8){\rm~GHz}$} and to adjust the detuning dynamically in a single run within a range of $0.7{\rm~GHz}$. Hence, the interaction strength can be adjusted  from  pulse to pulse to control the balance between signal strength and destructiveness (see Sec.~\ref{sec:SNR}).
The imaging light pulses are produced by an acousto-optic modulator, and are typically of 1~$\mu$s duration with a rectangular temporal envelope.

The experiments using DFFI are performed in the following sequence. Ultracold clouds of $^{87}$Rb atoms in the ${|f=2, m_f = 2 \rangle}$ state are produced by forced radio frequency evaporation in a Ioffe-Pritchard magnetic trap~\cite{Park2012} with typical axial and radial trapping frequencies of $\omega_z=2\pi\times17{\rm~ Hz}$ and \mbox{$\omega_r=2\pi\times192{\rm~ Hz}$} respectively. The temperature and number of atoms in the cloud are adjusted by tailoring the evaporation sequence. The Faraday imaging light propagates along the symmetry axis ($z$-axis in Fig.~\ref{fig:imaging_setup}) of the Ioffe-Pritchard trap; this corresponds to the direction of the trap's bias magnetic field.
At the end of each experimental sequence, we acquire a  time-of-flight absorption image. This provides an independent measurement of the number of atoms and the temperature of the cloud. 

\begin{figure}[t]
	\includegraphics[width=8.5cm]{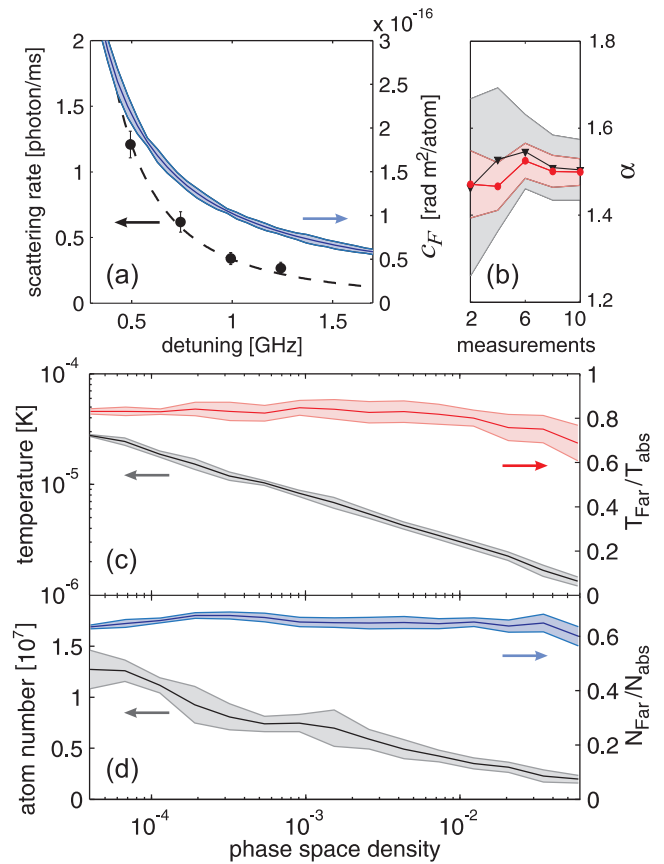}
  \caption{Characterization of DFFI. (a) Faraday coefficient $c_F$ and photon scattering rate as a function of the detuning $\Delta$. (b) Absorption image calibration factor $\alpha$ depending on the number of measurements with (dots) and without (triangles) Faraday scaling. (c), (d) Temperature and atom number obtained from DFFI compared to results from absorption imaging. Arrows in the figures indicate the appropriate axes. In (a), (c), and (d), the data has been binned  and the shaded regions indicate the standard error for each bin; in (b), the shaded region is the standard error of the measured value of $\alpha$ (see text). }
  \label{fig:detCompar_combined}
\end{figure}

It is of great importance to confirm that DFFI can precisely measure relevant properties of the atomic clouds. We first investigate the Faraday coefficient $c_F$ as a function of the laser detuning. Figure~\ref{fig:detCompar_combined}~(a) shows this dependence, where $c_F=\theta^\textrm{sum}/N_\textrm{abs}$ is obtained experimentally by summing $\theta_\mathrm{F}$ (obtained from Eq.~\eqref{sintheta}) over all CCD pixels and $N_\textrm{abs}$ is the atom number obtained from the absorption image. We focus on the characterization of the blue detuned side, to avoid complications arising from attractive  molecular resonances~\cite{Kaminski2012}.
In each experimental sequence, we prepared a thermal cloud at $3 {\rm~ \mu K}$ and took 35 Faraday images of $1 {\rm~ \mu s}$ duration separated by 4.7~ms at a pulse power of $160 {\rm~ \mu W}$; during the image acquisition, the laser detuning was swept over ${700 {\rm~ MHz}}$. The data agrees well with the theoretical expectation up to an overall scaling factor of 0.64. We ascribe this discrepancy to the spatial inhomogeneity in the magnetic potential and systematic calibration effects in absorption imaging~\cite{Yefsah2011}. Nonetheless, the agreement is good in light of previous work~\cite{Kubasik2009,Kaminski2012} and justifies neglecting the tensor terms in Eq.~\eqref{inter_ham} that would  induce detuning dependent corrections.

The destructiveness of DFFI is of similar importance. It was investigated at four different detunings by exposing the cloud to Faraday light for various durations and subsequently measuring the resulting temperature in absorption images. To obtain the scattering rate from the cloud temperature, we assume that each scattering event transfers twice the photon recoil energy and the heat capacity of an atom is $\frac{3}{2} k_B$. The measured scattering rate shown in Fig.~\ref{fig:detCompar_combined}~(a)
is consistent with this simple theoretical estimate to within a factor of two.
Figure~\ref{fig:detCompar_combined}~(a) also illustrates that the scattering rate decays as $1/\Delta^2$ whereas the Faraday coefficient falls off as $1/\Delta$. This well known difference is the key feature for the non-destructive character of the method.

To investigate the  measurement precision, we employ DFFI at different times in the evaporation sequence, sampling an atom number range from $1.5\times 10^7$ to $1.6\times 10^6$ and a temperature range from $30{\rm~ \mu K}$ to $1{\rm~ \mu K}$. The upper and lower limits of this range are set by the EMCCD size and the magnification of the detection system. In each experimental sequence, 20 Faraday images are taken, out of which the first six contain atoms and an average of the rest provides the intensity reference $I_0$. The pulse parameters correspond to those in Fig.~\ref{fig:detCompar_combined}~(a) at detuning +750~MHz, leading to an absorption probability per DFFI pulse of $6\times 10^{-4}$. The measured temperatures and atom numbers are shown in Fig.~\ref{fig:detCompar_combined}~(c) and (d) as a function of phase space density. In both cases DFFI allows for precise measurements over the entire parameter range, despite the fact that atom number and temperature are changed dramatically. To verify the accuracy of these measurements, we have taken a calibrated absorption image~\cite{Reinaudi2007} at the end of each experimental sequence. Figures~\ref{fig:detCompar_combined}~(c) and (d) show that the proportionality factor between the two methods is essentially constant. It is ${0.82\pm0.09}$ for the temperature, where the measurement uncertainty is the standard error of the entire data set. This represents good agreement considering the previously mentioned systematic effects in both methods. Consistent with the results in Fig.~\ref{fig:detCompar_combined}~(a), the atom number proportionality factor is ${0.65\pm0.03}$. This result confirms that DFFI provides precise non-destructive measurements and good accuracy can be obtained by appropriate scaling of the results.

\section{DFFI application examples}

\subsection{Circumventing shot-to-shot fluctuations}

Having established the functionality of the method, we turn to the first application of DFFI: We demonstrate the reduction of the deleterious effects of shot-to-shot fluctuations in an ultracold atomic sample. In general, creating a source of ultracold atoms with a reproducible atom number is a notoriously difficult experimental task. These shot-to-shot atom number fluctuations currently limit many fundamental investigations since the dynamics has to be averaged over many realizations. A number of previous experiments have implemented a correction for these fluctuations based on non-destructive measurements to study, e.g., inelastic collisions~\cite{Roberts2000}, photo-association~\cite{McKenzie2002}, and classical number fluctuations\cite{Sawyer2012a}.

We apply DFFI to provide a normalization procedure for the atom number calibration of absorption imaging. In the standard method~\cite{Reinaudi2007}, the scattering cross section is scaled by a factor $\alpha$. This factor is obtained by acquiring absorption images under identical conditions at various light intensities. Based on the expected functional dependence of the scattering cross section on the light intensity, $\alpha$ is chosen such that the dependence of the atom number on the light intensity is minimal. This method assumes identical atom numbers in subsequent experiments. Due to shot-to-shot fluctuations, a good estimate of $\alpha$ typically requires tens of measurements.

Here, we modify the procedure by adding a DFFI pulse in each experimental sequence. This allows us to perform the calibration procedure with a normalized atom number $N_\mathrm{abs}/N_\mathrm{DFFI}$. Figure~\ref{fig:detCompar_combined}~(b) shows $\alpha$ with and without the DFFI normalization as a function of the number of sampled light intensities.  The first point in Fig.~\ref{fig:detCompar_combined}~(b) corresponds to   two values of measured intensity. 
For each value of the  intensity, we performed three experimental runs. For the first point, $\alpha$ is given by the mean of the nine possible combinations, and the uncertainty is given by the standard deviation of this sample set.   Since normalization with DFFI avoids shot-to-shot fluctuations, a good estimate can be obtained with considerably fewer experimental points; in fact, two are sufficient in our case.

\subsection{Vector field magnetometry}\label{subsect:magnetometry}

Due to the magnetic field dependence of the Faraday effect, DFFI opens up new avenues in magnetometry. Vapor cell optical magnetometers~\cite{Budker2007} have been extremely successful, reaching sensitivities competitive with state-of-the-art SQUID magnetometers and allow for both spatial resolution~\cite{Kominis2003} and vector field magnetometry~\cite{Seltzer2004}.
Due to atomic motion, the spatial resolution is typically limited to millimeter length scales. On the other hand, ultracold atoms hold the promise for orders of magnitude higher precision due to the reduced thermal motion. Spatially resolved magnetometry has been realized in, e.g., dark optical tweezers~\cite{Fatemi2010} and Bose-Einstein condensates~\cite{Wildermuth2006,Vengalattore2007}, and  vector magnetometers based on cold atomic clouds were recently realized by two different methods~\cite{Smith2011,Koschorreck2011}.
To date, however, all realizations have been limited in interrogation time and in spatial resolution due to residual motion along a weakly confining trap axis.
In this work, we take an important conceptual step towards higher spatial resolution by realizing a single shot vector magnetometer based on ultracold atoms in an optical lattice. The method is an adaptation of a standard strategy of vapor cell magnetometers~\cite{Seltzer2004} relying on time dependent control of additional magnetic bias fields.
In principle, our approach allows for spatially resolved magnetometry down to the scale of a single lattice site ($\approx0.5~\mu$m).

\begin{figure}[t]
	\includegraphics[width=8.5cm]{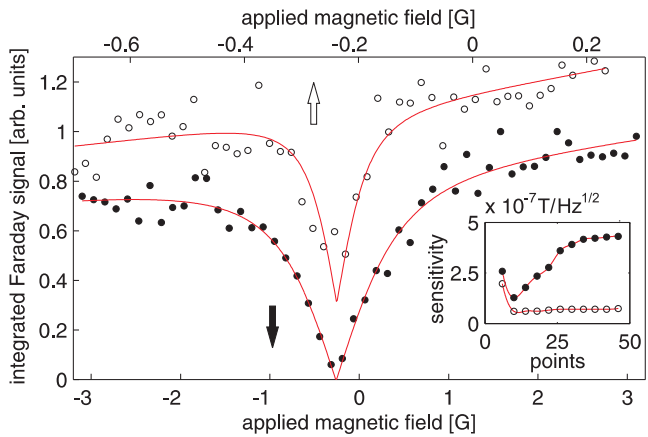}
  \caption{Single-run magnetometry in an optical lattice. DFFI signal as a function of the applied magnetic field along the $z$-axis. Open circles: Magnetic field sweep over 0.93~G at $|\mathbf{B_r}|=0.053$~G. Full dots: Magnetic field sweep over 6.2~G at $|\mathbf{B_r}|=1.03$~G. The inset shows the sensitivity of the offset field extraction for the two realizations as a function of number of data included points (centered around the signal minimum). The sensitivity is estimated as the error of the fit times the square root of time taken to record the included data points. }
	\label{fig:magnetometry}
\end{figure}

To realize this lattice magnetometer, the atomic cloud is  transferred into a 1D vertical lattice at a wavelength of 914~nm, whereupon we sweep the magnitude of an additional magnetic field applied along the $z$-axis. During this sweep, 50 DFFI pulses are taken to obtain the integrated Faraday signal at each applied magnetic field $B_z$  for two values of the transverse magnetic field, as shown in Fig.~\ref{fig:magnetometry}. The data was normalized and fitted with  $\rme^{-t/\tau} |B_z-B_{z_0}|/|\mathbf{B}|$, where the modulus is taken because our method is not sensitive to the sign of the Faraday rotation. The fit yields the offset field in the $z$-direction $B_{z_0}$ and the magnitude of the transverse field $|\mathbf{B_r}|$. The exponential factor in the fit function accounts for  atom loss during the sweep. In a first approach to quantify the sensitivity of such a time dependent vector magnetometer, the precision of extracting $B_{z_0}$ was evaluated as a function of the number of included data points (Fig.~\ref{fig:magnetometry} inset) yielding best values of $0.6 \times 10^{-7} ~{\rm T/\sqrt{Hz}}$ for the smaller sweep. The sweeps yield an offset field \mbox{$B_{z_0}=-0.252\pm0.013$~G}, which is in agreement with our microwave calibration. This demonstration shows new avenues for magnetometry with DFFI, which could be  exploited further with an optimized magnetometry sequence.

\subsection{Spatially resolved cloud dynamics}

Finally, DFFI permits the non-destructive investigation of spatial dynamics. We demonstrate this by monitoring the position of the atomic cloud as it oscillates in a harmonic potential. Since a single cloud can be probed repeatedly, one can map its trajectory in a single experimental run. Non-destructive imaging of oscillations  has previously been realized for a limited number of detection pulses~\cite{Stamper-Kurn1998}. In addition, continuous measurements without spatial resolution have been employed to monitor breathing~\cite{Petrov2007} and center-of-mass oscillations~\cite{Kohnen2011}; however, this approach fails for more complicated trajectories, e.g., when the position of the trapping potential is dynamically varied during the oscillation. 

\begin{figure}[]
	\includegraphics[width=8.5cm]{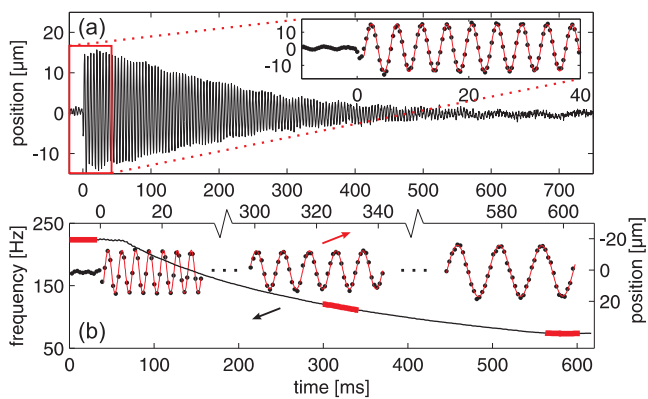}
	\caption{Monitoring of spatial dynamics. (a) Non-destructive measurement of the cloud position during a damped oscillation. (b) Non-destructive measurement of the cloud position during a decompression of the magnetic trap. The cloud position  and oscillation frequency are shown within three time intervals during the decompression.}
	\label{fig:osc_combined}
\end{figure}

Figure~\ref{fig:osc_combined}~(a) shows the position of the cloud recorded in a single experiment by acquiring a total of 2000 images at intervals of 0.402~ms. Initially, a cloud of about $10^6$ atoms at $1{\rm~ \mu K}$ was created in a magnetic trap and the imaging was started. Shortly afterwards ($t = 0$), the magnetic trap was turned off for a duration of ${\rm~ 70~\mu s}$, which initiated a strong vertical oscillation. The initial part of the oscillation was fitted to obtain a trapping frequency of ${222.44\pm0.06 {\rm~ Hz}}$ as shown in the inset of Fig.~\ref{fig:osc_combined}~(a). The residual anharmonicity of the trap makes the system ergodic and slowly transfers the collective motion of the atoms into thermal energy. This results in a decrease of the oscillation amplitude and heating of the cloud which can be observed simultaneously using DFFI.

This continuous probing of the spatial dynamics enables us to monitor dynamic changes of the system in a single sequence. To demonstrate this, we have observed an oscillating cloud during a decompression of the trap. Again, we prepared a cold cloud in a magnetic trap and started the acquisition of 750 images at intervals of 0.89~ms. The oscillations were initiated  at time $t = 0$ and after a hold time of 60~ms we began to decrease the current of the magnetic trap and simultaneously increase the bias field. The resulting decrease in the trapping frequency caused the cloud to sag due to gravity while it continued to oscillate. By subtracting the shift of the equilibrium position we obtained the chirped oscillations shown in Fig.~\ref{fig:osc_combined}~(b). The oscillations were fitted  within short time intervals to extract the time dependent trapping frequency. These measurements highlight the advantages of spatially resolved non-destructive probing, since the temperature and the in-trap equilibrium position cannot be extracted from non-spatially resolved  dispersive methods~\cite{Petrov2007, Kohnen2011}.

\section{signal-to-noise analysis of dispersive imaging techniques}
\label{sec:SNR}

When assessing the potential usefulness of an imaging technique, it is important to consider both the obtained signal strength and the associated noise.
In this section, we examine the signal-to-noise ratio of  four common dispersive imaging techniques.
To allow for a generalized treatment, we define a normalized signal function $S(\theta) \equiv I(\theta) / I_0$
where a general phase shift $\theta$ represents either $\theta_\mathrm{S}$ or $\theta_\mathrm{F}$ and  $I(\theta)$ and $I_0$ are the detected and the incident light intensities respectively.

Generally, the pixel values of the acquired images are subject to several sources of noise. The technical noise includes the readout noise of the camera, thermally induced dark counts, clock induced charges (CIC), and   noise
 from unstable light intensity or vibrations of the optical elements. All these noise contributions depend
on the particular implementation and can be minimized by careful design of the imaging setup and mitigated by post-processing~\cite{Ketterle_review}.
On the other hand, shot noise   cannot be eliminated and often becomes the dominant noise source. When calculating the SNR, we  therefore neglect technical noise sources and consider only the fundamental shot noise contribution.

The imaging regime where shot-noise predominates has  become   experimentally accessible within the last decade due to the introduction of Electron Multiplying Charge Coupled Device (EMCCD) Cameras~\cite{Robbins_IEEE_50_1227_2003,Hynecek2003}. Often, non destructive imaging requires very low light intensities such that the detected signal is comparable to the readout noise of the camera. An electron multiplying (EM) register before the readout register of the camera amplifies the signal and improves the sensitivity of an EMCCD camera to the level of single photon detection. The use of EM gain becomes profitable when the readout noise variance without EM gain is greater than the number of electrons $N_\mathrm{el}$ accumulated on a given pixel. The only disadvantage of the approach is that the random sequential character of the EM register amplifies any noise already present in the picture (such as the shot noise) by a factor of approximately $\sqrt{2}$.

For   dispersive imaging techniques, the number of electrons in each camera pixel can be calculated using \mbox{$N_{\rm el} = \eta N_{\rm ph,0}S(\theta)$,}
where $ N_{\rm ph,0}$ is the number of  incident photons per  pixel area.
The detection efficiency $\eta$ takes into account light losses in the imaging system as well as the quantum efficiency of the CCD.
Assuming  Poissonian statistics of the detected  light, the variance of a shot noise limited signal is $(\Delta N_{\rm el})^2 = N_{\rm el}$,
and the signal error is given by
\bn
\label{DeltaS}
  \Delta S = \left|\frac{dS}{dN_{\rm el}}\right|\Delta N_{\rm el} = \frac{\sqrt{\eta N_{\rm ph,0}S(\theta)}}{\eta N_{\rm ph,0}}
   =  \sqrt{\frac{S(\theta)}{\eta N_{\rm ph,0}}}.
\en
Hence, the EM   noise amplification by $\sqrt{2}$ effectively acts as a reduction of $\eta$ by a factor of 2.

To quantify the destructiveness of the imaging, we relate the number of incident photons to the photon absorption probability
\bn
  \label{Pabs}
   P_{\rm abs} \approx \frac{N_{\rm ph,0}}{A} \frac{\sigma_0}{\delta^2},
\en
where $A$ is the physical pixel area and $\sigma_0$ is the effective scattering cross-section ($\sigma_0 \approx \lambda^2/\pi$).
The above equation is valid in the large detuning limit $\delta \equiv \frac{\Delta}{\Gamma/2}\gg1$,
which is typically used to reduce diffraction effects.

Employing equation \eqref{Pabs}, the error in the measured phase shift becomes
\bn
  \label{DeltaTheta}
   \Delta \theta = \left|\frac{d\theta}{dS}\right| \Delta S
    =  \left|\frac{d\theta}{dS}\right| \frac{\sqrt{S(\theta)}}{|\delta|\sqrt{\eta P_{\rm abs} A/\sigma_0}}.
\en
Using the off-resonant scalar phase shift~\cite{Ketterle_review}
\begin{eqnarray}
  \label{thetaOffRes}
   \theta_\mathrm{S} = \frac{\sigma_0}{2\delta}\tilde{\rho},
\end{eqnarray}
where $\tilde{\rho}$ 
is the atomic column density (see Eq.~\eqref{far_rot}),
we can eliminate the explicit detuning dependence in Eq.~\eqref{DeltaTheta}, yielding
\bn
  \label{DeltaTheta2}
   \Delta \theta = \left|\frac{d\theta}{dS}\right| \frac{2|\theta_\mathrm{S}|\sqrt{S(\theta)}}{\tilde{\rho}\sqrt{ \eta \sigma_0 A P_{\rm abs}}}.
\en
The signal-to-noise ratio in the phase shift is thus given by
\begin{align}
\label{SNR_theta}
  {\rm SNR}_\theta &=  \frac{|\theta|}{\Delta \theta}=\frac{1}{2}\left|\frac{dS}{d\theta} \right| \frac{\Pi}{\sqrt{S(\theta)}},  \\
  \intertext{where  the scaling factor}
\Pi &\equiv \left| \frac{\theta}{\theta_\mathrm{S}} \right| \tilde{\rho}\sqrt{ \eta \sigma_0 A P_{\rm abs}}
\label{PiDef}
\end{align}
is a product of the phase shift ratio $|\theta /\theta_\mathrm{S}|$  and a term common to all dispersive methods.
Note that the signal-to-noise ratio is proportional to the measured atomic column density $\tilde{\rho}$, and that  it scales with the square root of photon absorption probability per atom $P_{\rm abs}$. Thus,  the absorption probability quantifies the trade-off between  measurement precision and destructiveness~\cite{PhysRevA.67.043609}.

The phase shift ratio $|\theta /\theta_\mathrm{S}|$ equals one for scalar imaging and is a non-trivial function of detuning for vector imaging. This function can be found from Hamiltonians \eqref{scal_ham} and \eqref{inter_ham} to be
\bn
\label{PhaseRatio}
   \frac{\theta_\mathrm{F}}{\theta_\mathrm{S}} = \frac{3 \langle f_z \rangle}{2}
\left[ \sum_{f^{\prime}} \frac{\alpha^{(\mathrm{vec})}_{f,f^{\prime}}}{\Delta_{f,f^{\prime}}} \right]
\left[ \sum_{f^{\prime}} \frac{\alpha^{(\mathrm{scal})}_{f,f^{\prime}}}{\Delta_{f,f^{\prime}}} \right]^{-1}.
\en
In table \ref{table:ratio}, we present the far-detuned limit of this ratio for the D transition in hydrogen like atoms with nuclear spin $I = 3/2$ such
as  $^{87}$Rb, $^{39}$K, $^{23}$Na, $^{7}$Li and also those with $I = 7/2$ such as $^{133}$Cs. In these cases, the ratio lies in the range of  0.25 to 1, resulting in a  reduced ${\rm SNR}_\theta$ for vector compared to scalar imaging methods.
In this far-detuned limit, the energy splitting between the hyperfine excited states can be neglected such that $\Delta_{f,f^{\prime}} \approx \Delta$ in Eq.~\eqref{PhaseRatio}.  Typically, we operate at detunings below this far-detuned limit so that  the full expression in Eq.~\eqref{PhaseRatio} is required to evaluate the phase-shift ratio. For example, a detuning of $\Delta = 2\pi \times 1$~GHz leads to $|\theta_\mathrm{F}/\theta_\mathrm{S}| = 0.59$. 
\begin{table}[t]
\caption{Far-detuned vector to scalar phase shift ratio $\left| \theta_\mathrm{F} / \theta_\mathrm{S} \right|$  for hydrogen-like atoms with nuclear spin $I$.}
\centering
\label{table:ratio}
   \begin{tabular}{ c c c | c c c }
 \hline \hline
      & \multicolumn{2}{c |}{$I = 3/2$} &  & \multicolumn{2}{c}{$I = 7/2$} \\
     $f$ &  ${\rm D}_1$ line  &  ${\rm D}_2$ line & $f$ & ${\rm D}_1$ line  &  ${\rm D}_2$ line  \\[0.5ex]
 \hline
      1 & 1/2 & 1/4 & 3 & 3/4 & 3/8 \\
      2 & 1    & 1/2 & 4 & 1    & 1/2 \\
\hline \hline
   \end{tabular}
\end{table}

In the following, we compare the different dispersive imaging techniques using this formalism. We neglect all experimental imperfections such as a non-ideal beam block in DFSI, incorrect phase-plate placement in PCI, and non-zero cube suppression in DFFI.
Provided that the measurement destructiveness $P_{\rm abs}$ is low,
the probe light transmission coefficient can be set to one and the detected light intensity for the two scalar imaging techniques
is given by~\cite{Ketterle_review}
\begin{eqnarray}
 I^{\rm (DFSI)} &=& I_0 \left[ 2 - 2\cos(\theta_\mathrm{S})  \right], \\
 I^{\rm (PCI)} &=& I_0 \left[ 3 - \sqrt{8} \cos(\theta_\mathrm{S} + \pi/4) \right] \\
&=&I_0 \left[ 3 - 2\right( \cos(\theta_\mathrm{S} ) - \sin(\theta_\mathrm{S} ) \left)  \right],
\end{eqnarray}
where  a phase shift of $\pi/2$  relative to the unscattered component was included in PCI. The intensity obtained for the DFFI method was given in Eq.~\eqref{eqn:thetaF}.
Finally, in the DPFI method, the probe light is initially polarized at $45^\circ$ with respect to the PBS axis. The horizontally and vertically
polarized components are imaged separately on a camera according to
\begin{eqnarray}
 I^{\rm (DPFI)}_H &=& I_0 \left[ 1 + \sin(2\theta_\mathrm{F})  \right]/2, \\
 I^{\rm (DPFI)}_V &=& I_0 \left[ 1 - \sin(2\theta_\mathrm{F})  \right]/2.
\end{eqnarray}
The signal in DPFI is obtained by subtracting the two images. This has the advantage that common-mode noise is rejected and that a full measurement of the Stokes vectors $S_0\propto I_H+ I_V$ and $S_1\propto I_H- I_V$ is possible. However, the variance of the resulting signal is proportional to the sum of the variances in each image, making the signal error independent of the rotation angle.
Consequently, for the DPFI method, Eq.~\eqref{DeltaS} must be replaced by $\Delta S = 1/\sqrt{\eta N_{\rm ph,0}}$. 
Note, that at $\theta_\mathrm{S} = 0$, the signal error in the PCI method has the same value.

\begin{table}[t]
\caption{Signal properties of common dispersive imaging methods.}
\centering
\label{table:properties}
   \begin{tabular}{ c  c c }
 \hline \hline
     Method & $S(\theta)$ &  ${\rm SNR}_\theta/\Pi$ \\
 \hline
   DFSI &  $2 - 2\cos(\theta_\mathrm{S})$                                                                        & $\left| \cos(\theta_\mathrm{S}/2) \right|$   \\
   PCI   & $3 - 2( \cos(\theta_\mathrm{S} )-\sin(\theta_\mathrm{S} ))$    &
  $\frac{\left| \sin(\theta_\mathrm{S} )+\cos(\theta_\mathrm{S} ) \right|}{\sqrt{3 - 2( \cos(\theta_\mathrm{S} )-\sin(\theta_\mathrm{S} ))}}$  \\
   DFFI & $\sin^2(\theta_\mathrm{F})$                                                                             & $| \cos(\theta_\mathrm{F})|$  \\
   DPFI & $\sin(2\theta_\mathrm{F})$                                                                       & $| \cos(2\theta_\mathrm{F})|$  \\
\hline \hline
   \end{tabular}
\end{table}

Table \ref{table:properties} summarizes the signal function $S(\theta)$ and the resulting ${\rm SNR}_\theta$ for the four dispersive techniques.
For small angles, the signal $S(\theta)$  grows quadratically with the acquired angle  in the two dark field techniques DFSI and DFFI, whereas  it grows linearly in PCI and DPFI. For this reason,  the last two  are often considered superior for  non-destructive measurements~\cite{Ketterle_review}.  
However, a disadvantage of these methods is that, for  small angles, all of the light is sent onto the camera and  results in a constant shot noise contribution.
 This is not the case for the dark field methods, where only the scattered light is collected, thus reducing the shot noise contribution at low angles. This leads to the unexpected result ${\rm SNR}_\theta=\Pi$ at $\theta=0$ for all four methods.
In this limit, the signal to noise ratios of the four methods differ only  by the ratio $|\theta_\mathrm{F} /\theta_\mathrm{S}|$.

 To clarify the unexpected result that the signal-to-noise ratio is maximized for zero signal, we briefly discuss the derivation. 
 The  result arises from the interdependence of the quantities we have used to present a conceptually unified treatment of the four dispersive imaging methods.
The key feature is the scaling of the ${\rm SNR}_\theta$ with  $\Pi$: the scaling factor $\Pi$ represents a set of experimental conditions that is determined by the choice of atom-light interaction (scalar or vector), the atomic density, the imaging system and the  destructiveness.
However, fixing those quantities does not impose any constraints on the choice of detuning $\delta$, as long as the approximations 
 $P_{\rm abs}\propto N_{\rm ph,0}/\delta^2$ and $\theta\propto 1/\delta$ remain valid. 
Thus, a reduction in the phase shift ($\theta \rightarrow 0$) caused by an increase in the detuning $\delta$
can always be compensated by an increase in the number of photons $N_{\rm ph,0}$, preserving the value of ${\rm SNR}_\theta$ due to the reduced relevance of the shot noise---the only source of noise in this model.
In practice, however, shot noise limited detection is only feasible for a limited number of photons due to technical limitations such as the  finite dynamic range of the detector. To place this in experimental context, 
although  ${\rm SNR}_\theta$ is maximized at $\theta = 0$, 
 all four methods reach a sensitivity of more than 99\% of the maximum 
 for a four degree phase shift, implying that the signal-to-noise ratio saturates as $\delta \rightarrow \infty$.
\begin{figure}[t]
	\includegraphics[width=8.5cm]{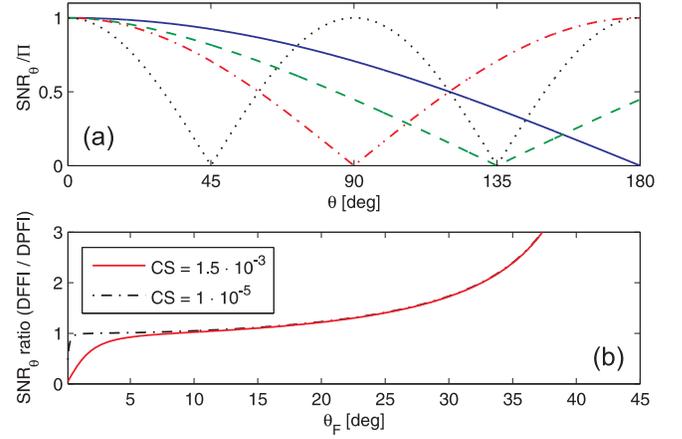}
	\caption{Signal-to-noise ratio in the measured phase shift. (a) ${\rm SNR}_\theta /\Pi$ for common dispersive imaging techniques (DFSI blue solid; PCI green dashed;
         DFFI red dot-dashed; DPFI black dotted) as a function of the respective phase shift. (b) Relative ${\rm SNR}_\theta$ ratio for DFFI and DPFI vs. Faraday rotation angle for two values of cube suppression $\CS$. }
	\label{fig:SNRtheta}
\end{figure}

To  illustrate the signal-to-noise behavior of the four dispersive imaging methods, Fig.~\ref{fig:SNRtheta}~(a) shows  ${\rm SNR}_\theta/\Pi$ versus  phase shift $\theta$.
At low angles, all methods yield a quadratic decrease in ${\rm SNR}_\theta/\Pi$. As the angle is increased, the signal function for each method reaches a local maximum. This maximum corresponds to a  zero in the derivative of the signal function and  leads to the sequence of  zeros in ${\rm SNR}_\theta/\Pi$ shown in Fig.~\ref{fig:SNRtheta}~(a). The DPFI technique, oscillating most rapidly in the ${\rm SNR}_\theta$, reaches its first zero at $\theta=45^{\circ}$, where the full intensity is sent onto a single port. An equivalent situation occurs in DFFI, when the phase reaches $90^{\circ}$.
The maxima in the scalar methods can be understood by considering the phasor diagram of each method (see, e.g., Fig.~8 in Ref.~\cite{Ketterle_review}). In PCI, the unscattered light is shifted by $90^{\circ}$ and added to the scattered field. The intensity of this sum reaches a maximum at $\theta=135^{\circ}$.
Finally, in DFSI the  signal corresponds to the difference between the incident and scattered electric fields. At $\theta=180^{\circ}$ the electric field vector of the scattered light is anti-parallel to the incident electric field resulting in the maximal scattered fraction. 

Figure \ref{fig:SNRtheta}~(b) shows a comparison of the two Faraday imaging techniques in a more realistic model that includes the cube suppression. We present the result for our experimental value of $\CS = 1.5\times10^{-3}$  and for a high-quality (Glan-Thompson) polarizer with $\CS = 10^{-5}$. For low angles, DPFI has a superior signal-to-noise ratio to that of DFFI, although this advantage is negligible when using a high-quality polarizer. For the  typical rotation angles used in our experiment $5^\circ\lesssim\theta \leq 45^\circ$, DFFI outperforms DPFI.  

To conclude this analysis, all methods yield a signal-to-noise ratio ${\rm SNR}_\theta/\Pi$ that is  periodic in the acquired phase and all attain the same maximum value. Therefore, each method has a certain range within which it is superior to the others, making the choice of method a matter of experimental convenience. A similar conclusion is reached when considering the broader choice between scalar and vector methods. 
 While the scalar phase shift is typically  larger than that generated by vector methods,  the sensitivity of the vector methods to the magnetic substate is the key difference. Thus, the preferred method is determined by the application. For instance,  the sensitivity to magnetic substate is of particular importance in magnetometry (see section \ref{subsect:magnetometry}) and the imaging of multi-component quantum gases~\cite{Vengalattore2007}.
 
\section{Conclusion}
We have characterized a simple and precise method to non-destructively probe ultracold atomic samples. This dark field Faraday imaging  method can be implemented by inserting a single polarizer into a standard absorption imaging system. We have investigated DFFI over a wide range of parameters and have  shown that it provides precise measurements of the atomic temperature and density. The signal-to-noise ratio of the method was shown to be  similar to other conventional dispersive imaging techniques. The potential of the method as a tool for the characterization and manipulation of ultracold atoms was demonstrated in three applications: shot-to-shot atom number fluctuation compensation in the calibration of an absorption imaging system, single-shot spatially resolved vector magnetometry in an optical lattice, and  non-destructive imaging of spatial dynamics. In combination with fast data analysis, this method will allow for quantum engineering using measurements and feedback. In particular, it will be interesting to investigate if it can extend the methods for short-cuts to adiabaticity currently under investigation~\cite{Schaff2011} and if it allows for the deterministic production of exotic quantum states.

We thank J.~H.~M\"uller for fruitful discussions. We acknowledge support from the Danish National Research Foundation, the Danish Council for Independent Research, and the Lundbeck Foundation.
%\bibliography{gajdacz}

\end{document}